\documentclass[final,5p,times,twocolumn]{elsarticle}

\usepackage{graphicx}
\usepackage{amssymb}
\usepackage{amsmath}
\journal{Physics Letters B}
\newcommand{\bea}{\begin{eqnarray}}
\newcommand{\eea}{\end{eqnarray}}
\newcommand{\bi}{\begin{itemize}}
\newcommand{\ei}{\end{itemize}}
\newcommand{\benu}{\begin{enumerate}}
\newcommand{\eenu}{\end{enumerate}}
\newcommand{\nn}{\nonumber}
\newfont{\bg}{cmr10 scaled\magstep4}
\newcommand{\bigzerol}{\smash{\hbox{\bg 0}}}
\def\Cv{\mbox{\boldmath $C$}}

\def\Kv{\mbox{\boldmath $K$}}

\def\qv{\mbox{\boldmath $q$}}
\def\kv{\mbox{\boldmath $k$}}

\begin{document}

\begin{frontmatter}



%--- title
\title{Photon Structure Function in Supersymmetric QCD Revisited}


%--- author informations 

\author[kyoto]{Ryo Sahara}
\ead{sahara@scphys.kyoto-u.ac.jp}

\author[kyoto]{Tsuneo Uematsu}
\ead{uematsu@scphys.kyoto-u.ac.jp}

\author[IPAST]{Yoshio Kitadono}
\ead{kitadono@phys.sinica.edu.tw}

%--- institute information

\address[kyoto]{Department of Physics, 
 Graduate School of Science, Kyoto University,\\
    Kitashirakawa, Kyoto 606-8502, Japan}
%\address[MISC]{Maskawa Institute for Science and Culture, 
%Kyoto Sangyo University, Kyoto 603-8555, Japan}
\address[IPAST]{Institute of Physics, Academia Sinica, Taipei, Taiwan}

%--- abstract
\begin{abstract}
We investigate the virtual photon structure function in the 
supersymmetric QCD (SQCD), where we have squarks and gluinos 
in addition to the quarks and gluons. 
Taking into account the heavy particle mass effects to 
the leading order in QCD and SQCD we evaluate the photon structure function
and numerically study its behavior for the QCD and SQCD cases.
\end{abstract}

%--- keywords

\end{frontmatter}

% \linenumbers

%\tableofcontents
%--------- main text --------------\
%\section{Introduction \label{introduction}}
Since the experiments at the Large Hadron Collider (LHC) \cite{LHC} 
started there has been much anticipation for the signals 
of the Higgs boson as well as for an evidence of the new physics 
beyond Standard Model such as supersymmetry (SUSY). 
Once these signals are observed more precise measurement needs 
to be carried out at the future $e^{+}e^{-}$ collider, so called 
International Linear Collider (ILC) \cite{ILC}.  
In such a case, it is important to know the theoretical 
predictions at high energies based on QCD.

It is well known that, in $e^+ e^-$ collision experiments, the cross section
for the two-photon processes 
$e^+ e^- \rightarrow e^+ e^- + {\rm hadrons}$
dominates at high energies over 
the one-photon annihilation process~\cite{twophoton}.
%$e^+ e^- \rightarrow \gamma^*
%\rightarrow {\rm hadrons}$
%~\cite{twophoton}. 
We consider here the two-photon processes in the
double-tag events where both of the outgoing $e^+$ and $e^-$ are 
detected.
Especially, the case in which one of the virtual photon
is far off-shell (large $Q^2\equiv -q^2$), while the other is close to
%the mass-shell (small $P^2=-p^2$), can be viewed as a
the mass-shell (small $P^2=-p^2$), with
$\Lambda^2\ll P^2\ll Q^2$ ($\Lambda$: QCD scale parameter), 
can be viewed as a deep-inelastic 
scattering where the target is a virtual photon 
and we can calculate the photon structure 
functions in perturbation theories
~\cite{Review,QPM,Witten,BB,Altarelli,Dewitt,GR1983,UW,Rossi}.
%%%%%%%%%%%%%%%%%%%%%%%%%%%%%%%%%%%%%
\begin{figure}[hbt]
\begin{center}
\includegraphics[scale=0.25]{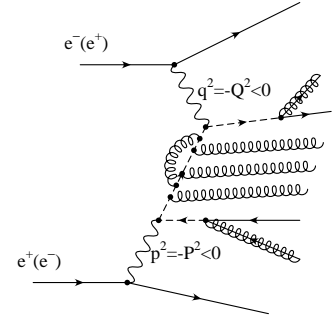}
\vspace{-0.3cm}
\caption{\label{super-2photon} $e^+~e^-$ two-photon processes in 
supersymmetric QCD. The solid (dashed) line denotes the quark (squark),
while the spiral (spiral-straight) line implies the gluon (gluino).
} 
\end{center}
\end{figure}
%%%%%%%%%%%%%%%%%%%%%%%%%%%%%%%%%%%%%

%Many authors studied the heavy quark mass effects 
%in the photon structure functions~\cite{GRSch,SSU,CJKL}.
%The heavy quark mass effects in the photon structure functions were
%studied in the literature~\cite{GRSch,SSU,CJKL}.
%nucleon structure functions
%were investigated in the framework of
%operator product expansion or QCD improved parton model approach.
%We particulary consider the virtual photon target in the
%kinematic region, $\Lambda^2\ll P^2\ll Q^2$, where we can perturbatively
%calculate the whole structure function.
Some time ago 
the effects of supersymmetry on two-photon process were studied in the 
literature 
\cite{Reya,Ross-Weston,Drees-Gluck-Reya,Antoniadis-Kounnas-Lacaze}. 
In this paper based on the framework of treating heavy parton 
distributions \cite{KSUU,KSUUResum} 
we reexamine the effects of the squarks and gluinos
appearing in SUSY QCD (SQCD) on the photon structure functions to the leading
order in SQCD 
which can be 
measured in the two-photon processes of $e^+e^-$ collision illustrated
in Fig.\ref{super-2photon}. 
%%%%%%%%%%%%%%%%%%%%%%%%%%%%%%%%%%%%%%%%%%%%%%%
\section{Evolution equations for the SUSY QCD}
%%%%%%%%%%%%%%%%%%%%%%%%%%%%%%%%%%%%%%%%%%%%%%%

We consider the DGLAP type evolution equations for the parton
distribution functions inside the virtual photon with the mass squared, 
$P^2$, in SQCD where we have squarks
and gluinos in addition to the ordinary quarks and gluons. 
Evolution equation to the leading order (LO) 
in SQCD reads as in QCD \cite{Sasaki-Uematsu}:
\bea
\frac{d\qv^\gamma(t)}{dt}=
\qv^\gamma(t)\otimes 
P^{(0)}+\frac{\alpha}{\alpha_s(t)}\kv^{(0)}~,\label{evo-lo}
\eea
where $P^{(0)}$ and $\kv^{(0)}$ are 1-loop parton-parton and photon-parton
splitting functions, respectively (see Appendix). 
The symbol $\otimes$ denotes the convolution
between the splitting function and the parton distribution function.
 The variable $t$ is defined in terms of the running 
coupling $\alpha_s$ as \cite{Furmanski-Petronzio}:
\bea
\hspace{-0.5cm}
t=\frac{2}{\beta_0}\ln\frac{\alpha_s(P^2)}{\alpha_s(Q^2)},\quad 
%\frac{d\alpha_s}{dt}=-\frac{\beta_0}{2}\alpha_s,\quad 
%\frac{\alpha_s(t)}{\alpha_s(0)}=e^{-\frac{\beta_0}{2}t}~.
\frac{d\alpha_s(Q^2)}{d\ln{Q^2}}=-\beta_0\frac{\alpha_s(Q^2)^2}{4\pi}
+{\cal O}(\alpha_s(Q^2)^3)
\label{evo-val}
\eea
%where $d\alpha_s/dt=-(\beta_0/2)\alpha_s$ and
with the parton distributions probed by the virtual photon with 
mass squared $Q^2$ as
\bea
\qv^\gamma(t)=(G,\lambda,q_1, \cdots q_{n_f}, s_1, \cdots, s_{n_f}),
\eea
where $n_f$ is the number of active flavors. In eq.(\ref{evo-val}),
$\beta_0=9-n_f$ for SQCD.
We denote the distribution function of the 
$i$-th flavor quark, squark by $q_i(x,Q^2,P^2)$, 
$s_i(x,Q^2,P^2)$, ($i=1,\cdots, n_{f}$),  and the gluon, gluino 
by $G(x,Q^2,P^2)$, $\lambda(x,Q^2,P^2)$, respectively. 
The 1-loop splitting functions were obtained in 
\cite{Kounnas-Ross,Jones-Smith}.
We first consider the case where all the particles are massless.
Although this is an unrealistic case, it is instructive to consider
the massless case for the later treatment of the realistic case with
the heavy mass effects. 

For the massless partons the evolution starts at $Q^2=P^2$ and
hence we have the initial condition $\qv^\gamma(t=0)=0~$\cite{UW}.

The 1-loop splitting function is given by (see Appendix A)
\bea
P^{(0)}=
\left(
\begin{array}{cccc}
P_{GG}&P_{\lambda G}&P_{qG}&P_{sG}\\
P_{G\lambda}&P_{\lambda\lambda}&P_{q\lambda}&P_{s\lambda}\\
P_{Gq}&P_{\lambda q}&P_{qq}&P_{sq}\\
P_{Gs}&P_{\lambda s}&P_{qs}&P_{ss}
\end{array}
\right)~,
\eea
where $P_{AB}$ is a splitting function of $B$ parton to $A$ parton
with $A,B=G,\lambda,q$ and $s$.
While the splitting functions of the photon into the partons 
$G,\lambda,q$ and $s$, are denoted as (see Appendix B)
\bea
\kv^{(0)}=(k_G,k_\lambda,k_q,k_s)~.
\eea

We introduce the flavor-nonsinglet (NS) combinations
of the quark and squark distribution functions as
\bea
&&q_{NS}(x,Q^2,P^2)=\sum_{i=1}^{n_f}(e_i^2-\langle e^2\rangle)
q_i(x,Q^2,P^2)~,\\
&&s_{NS}(x,Q^2,P^2)=\sum_{i=1}^{n_f}(e_i^2-\langle e^2\rangle)
s_i(x,Q^2,P^2)~,
\eea
where $e_i$ is the $i$-th flavor charge and
$\langle e^2\rangle=\sum_i e_i^2/n_f$  is the average charge squared.
We also define the flavor-singlet (S) combinations for quarks and squarks
\bea
&&\Sigma(x,Q^2,P^2)=\sum_{i=1}^{n_f}q_i(x,Q^2,P^2)~,\\
&&S(x,Q^2,P^2)=\sum_{i=1}^{n_f}s_i(x,Q^2,P^2)~.
\eea
We now rearrange the parton components of $\qv^\gamma(t)$ using the
above flavor non-singlet and singlet combinations as:
\bea
\qv^\gamma(t)=(G,\lambda,\Sigma,S,q_{NS},s_{NS})~.
\eea
Then we have the following splitting function
\bea
P^{(0)}=
\left(
\begin{array}{cccc|cc}
P_{GG}&P_{\lambda G}&P_{qG}&P_{sG}& & \\
P_{G\lambda}&P_{\lambda\lambda}&P_{q\lambda}&P_{s\lambda}& &\bigzerol \\
P_{Gq}&P_{\lambda q}&P_{qq}&P_{sq}& & \\
P_{Gs}&P_{\lambda s}&P_{qs}&P_{ss}& & \\
\hline
& & & &P_{qq}&P_{sq}\\
& \bigzerol& & &P_{qs}&P_{ss}
\end{array}
\right)~.
\eea
Thus for the flavor-nonsinglet parton distributions
\bea
\qv_{NS}^\gamma=(q_{NS},s_{NS})~,
\eea
satisfy the following evolution equation:
\bea
\frac{d\qv_{NS}^\gamma}{dt}=\qv_{NS}^\gamma\otimes P^{(0)}_{NS}+
\frac{\alpha}{\alpha_s(t)}\kv^{(NS)}~,
\eea
where the splitting functions are
\bea
&&\hspace{-1cm}P^{(0)}_{NS}=
\left(
\begin{array}{cc}
P_{qq}&P_{sq} \\
P_{qs}&P_{ss} 
\end{array}
\right),\
\kv^{(NS)}=(K_q^{(NS)},K_s^{(NS)}),\\
&&\hspace{-1cm}K_q^{(NS)}=\sum_{i=1}^{n_f}(e_i^2-\langle e^2\rangle)k_{q_i},\
K_s^{(NS)}=\sum_{i=1}^{n_f}(e_i^2-\langle e^2\rangle)k_{s_i}~.
\eea
For the flavor-singlet parton distribution
\bea
\qv_{S}^\gamma=(G,\lambda,\Sigma,S)~,
\eea
we have
\bea
\frac{d\qv_{S}^\gamma}{dt}=\qv_{S}^\gamma\otimes P^{(0)}_{S}+
\frac{\alpha}{\alpha_s(t)}\kv^{(S)}~,
\eea
where
\bea
&&\hspace{-1cm}
P^{(0)}_{S}=
\left(
\begin{array}{cccc}
P_{GG}&P_{\lambda G}&P_{qG}&P_{sG} \\
P_{G\lambda}&P_{\lambda\lambda}&P_{q\lambda}&P_{s\lambda}\\
P_{Gq}&P_{\lambda q}&P_{qq}&P_{sq} \\
P_{Gs}&P_{\lambda s}&P_{qs}&P_{ss}
\end{array}
\right)~,\\ 
&&\hspace{-1cm}
\kv^{(S)}=(K_q^{(S)},K_s^{(S)}),\
K_q^{(S)}=\sum_{i=1}^{n_f}k_{q_i},\
K_s^{(S)}=\sum_{i=1}^{n_f}k_{s_i}.
\eea

Now we should notice that there exist the following supersymmetric
relations for the splitting functions \cite{Kounnas-Ross}:
\bea
&& P_{qq}+P_{sq}=P_{qs}+P_{ss}\equiv P_{\phi\phi},\\
&& P_{qG}+P_{sG}=P_{q\lambda}+P_{s\lambda}\equiv P_{\phi V},\\
&& P_{Gq}+P_{\lambda q}=P_{Gs}+P_{\lambda s}\equiv P_{V\phi},\\
&& P_{GG}+P_{\lambda G}=P_{G\lambda}+P_{\lambda\lambda}\equiv
P_{VV}~.
\eea
Hence if we introduce the following combinations
\bea
\hspace{1cm} \phi\equiv \Sigma+S,\quad V\equiv G+\lambda,
\eea
then we obtain the following compact form for the
mixing of the flavor-singlet part:
\bea
&&\frac{d\phi}{dt}=P_{\phi\phi}\otimes\phi
+P_{\phi V}\otimes V +\frac{\alpha}{\alpha_s(t)}K_\phi^{(S)},\\
&&\frac{dV}{dt}=P_{V\phi}\otimes\phi+ P_{V V}\otimes V,
\eea
where we denote $K_\phi^{(S)}=K_q^{(S)}+K_s^{(S)}$ 
and the flavor-nonsinglet part $\phi_{NS}=q_{NS}+s_{NS}$ becomes
\bea
\frac{d\phi_{NS}}{dt}=P_{\phi\phi}\otimes\phi_{NS}
+\frac{\alpha}{\alpha_s(t)}K_\phi^{(NS)},
\eea
where we introduced $K_\phi^{(NS)}=K_q^{(NS)}+K_s^{(NS)}$. 
In terms of the flavor singlet and non-singlet parton
distribution functions we can express the virtual photon
structure function $F_2^\gamma$ as
\bea
&&\hspace{-1cm}F_2^\gamma(x,Q^2,P^2)=x\sum_i e_i^2 (q_i(x,Q^2,P^2)
+s_i(x,Q^2,P^2))\nn\\
&=& x\sum_i (e_i^2-\langle e^2\rangle)(q_i(x,Q^2,P^2)
+s_i(x,Q^2,P^2))\nn\\
&&+x\langle e^2\rangle\sum_i(q_i(x,Q^2,P^2)
+s_i(x,Q^2,P^2))\nn\\
&=& x(q_{NS}(x,Q^2,P^2)+s_{NS}(x,Q^2,P^2))\nn\\
&&+x\langle e^2\rangle(\Sigma(x,Q^2,P^2)+S(x,Q^2,P^2))\nn\\
&=&x\phi_{NS}(x,Q^2,P^2)+x\langle e^2\rangle
\phi(x,Q^2,P^2)
\eea

In Fig.2 we have plotted the virtual photon structure
function $F_2^\gamma(x,Q^2,P^2)$ in the SQCD
as well as in the ordinary QCD
for $Q^2=(1000)^2$GeV$^2$ and $P^2=(10)^2$GeV$^2$.
We have also shown the quark as well as the squark components
of the virtual photon structure function $F_2^\gamma$ 
in the case of the SQCD. 
In contrast to the QCD, the momentum fraction carried
by the quarks in the SQCD case decreases due to the emission of 
the squarks and the gluinos. 
Hence the $x$-distribution of the quarks 
for the SQCD increases at small-$x$ and decreases at large $x$, i.e.
it becomes more flat compared to the QCD case as seen from 
the Fig.2. Adding the two components together we get
the $F_2^\gamma$ structure function for the SQCD which shows
a behavior quite different from that of the QCD.

%%%%%%%%%%%%%%%%%%%%%%%%%%%%%
\begin{figure}[hbt]
\begin{center}
\includegraphics[scale=0.62]{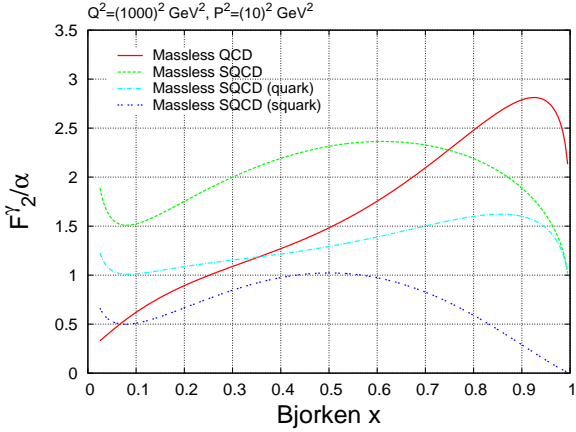}
\caption{
The virtual photon structure function $F_2^\gamma(x,Q^2,P^2)$ 
divided by the QED coupling constant
$\alpha$ for massless QCD (solid line) and SQCD (dashed line)
with $n_f=6$, $Q^2=(1000)^2$GeV$^2$ and $P^2=(10)^2$GeV$^2$. 
Also shown are the quark (dash-dotted line) and the
squark (double-dotted line) components.}
\end{center}
\end{figure}
\vspace{-1cm}
%%%%%%%%%%%%%%%%%%%%%%%%%%%%%%%%%%%%%
\section{Heavy parton mass effects}
%%%%%%%%%%%%%%%%%%%%%%%%%%%%%%%%%%%%%
Many authors have studied heavy quark mass effects in the nucleon
\cite{Buza-etal} and the photon structure functions~\cite{GRSch,SSU,CJKL}.
Now we consider the heavy parton mass effects, and we decompose
the parton distributions in the case where we have $n_f-1$ light quarks
and one heavy quark flavor which we take the $n_f$-th quark and 
all the squarks have the same heavy mass, while the gluino has
another heavy mass \cite{KSUU,KSUUResum}:
\bea
\qv^\gamma(t)=(G,\lambda,q_1,\cdots,q_{n_f-1}, s_1, \cdots, 
s_{n_f-1},q_H,s_H)~.
\eea
We denote the $i$-th light flavor quark, squark by $q_i(x,Q^2,P^2)$, 
$s_i(x,Q^2,P^2)$, ($i=1,\cdots, n_f-1$), one heavy quark and its superpartner
(squark) by $q_H$, $s_H$  and the gluon, gluino 
by $G(x,Q^2,P^2)$, $\lambda(x,Q^2,P^2)$, respectively. 

We now define light flavor-nonsinglet (LNS) and singlet (LS)
combination of the quark and the squark as follows:
\bea
&&\hspace{-1cm}q_{LNS}=\sum_{i=1}^{n_f-1}
\left(e_i^2-\langle e^2\rangle_L\right)q_i,\quad
s_{LNS}=\sum_{i=1}^{n_f-1}
\left(e_i^2-\langle e^2\rangle_L\right)s_i~,\nn\\
&&\hspace{-1cm}q_{LS}=\sum_{i=1}^{n_f-1}q_i,\quad s_{LS}=\sum_{i=1}^{n_f-1}s_i,
\quad \langle e^2\rangle_L=\frac{1}{n_f-1}\sum_{i=1}^{n_f-1}
e_i^2~.
\eea
Then we rearrange the parton distributions as
\bea
\qv^\gamma(t)=(G,\lambda,q_{LS},s_{LS},q_H,s_H,q_{LNS},s_{LNS})~.
\eea
The evolution equations and the splitting functions read
\bea
&&\hspace{-1cm}
\frac{d\qv^\gamma(t)}{dt}=
\qv^\gamma(t)\otimes 
P^{(0)}+\frac{\alpha}{\alpha_s(t)}\kv^{(0)}~,\quad
P^{(0)}=\left(
\begin{array}{c|c}
P^{LS}&0\\
\hline
0&P^{LNS}
\end{array}
\right)~,\nn\\
&&\hspace{-1cm}
P^{LS}\equiv
\left(
\begin{array}{cccccc}
P_{GG}&P_{\lambda G}&\frac{n_f-1}{n_f}P_{qG}
&\frac{n_f-1}{n_f}P_{sG}&\frac{1}{n_f}P_{qG}&\frac{1}{n_f}P_{sG}\\
P_{G\lambda}&P_{\lambda\lambda}&\frac{n_f-1}{n_f}P_{q\lambda}
&\frac{n_f-1}{n_f}P_{s\lambda}&\frac{1}{n_f}P_{q\lambda}&
\frac{1}{n_f}P_{s\lambda}\\
P_{Gq}&P_{\lambda q}&P_{qq}&P_{sq}&0&0\\
P_{Gs}&P_{\lambda s}&P_{qs}&P_{ss}&0&0\\
P_{Gq}&P_{\lambda q}&0&0&P_{qq}&P_{sq}\\
P_{Gs}&P_{\lambda s}&0&0&P_{qs}&P_{ss}
\end{array}
\right),\nn\\ 
&&\hspace{-1cm}
P^{LNS}\equiv
\left(
\begin{array}{cc}
P_{qq}&P_{sq}\\
P_{qs}&P_{ss}
\end{array}
\right)~.
\eea
While the photon-parton splitting functions are
\bea
\kv^{(0)}=(k_G,k_\lambda,k_{q_{LS}},k_{s_{LS}},k_{q_H},k_{s_H},k_{q_{LNS}},k_{s_{LNS}})
~.
\eea
Now we take into account the heavy mass effects by setting
the initial conditions for the heavy parton distribution
functions as discussed in \cite{KSUUResum,Fontannaz,AFG}.

We note here that 
the structure function $F_2^\gamma$ can be written as a convolution
of the parton distribution $\qv^\gamma(x,Q^2,P^2)$ and the Wilson
coefficient function $\Cv(x,Q^2)$:
\bea
F_2^\gamma(x,Q^2,P^2)/x=\qv^\gamma\otimes\Cv~.
\eea
The moments of the parton distributions are defined as
\bea
\qv^\gamma(n,t)\equiv \int_0^1dx x^{n-1}\qv^\gamma(x,Q^2,P^2)~,
\eea
where we put the initial conditions:
\bea
\qv^\gamma(n,t=0)=\left(0,\hat{\lambda}(n),0,\hat{s}_{LS}(n),\hat{q}_H(n),
\hat{s}_H(n),0,\hat{s}_{LNS}(n)\right)~,
\eea
and require that the following boundary conditions are
satisfied:
\bea
&&\hspace{-1cm}
\lambda(n,Q^2=m_\lambda^2)=0,\
s_{LS}(n,Q^2=m_{sq}^2)=0,\
q_H(n,Q^2=m_{H}^2)=0,\nn\\
&&\hspace{-1cm}
s_H(n,Q^2=m_{sq}^2)=0,\
s_{LNS}(n,Q^2=m_{sq}^2)=0,
\eea
where $m_\lambda$, $m_{sq}$ and $m_H$ are the mass of
the gluino, squarks and the heavy (here we take top) quark,
respectively. Note that here we take all the squarks have the same
mass $m_{sq}$.

By solving the evolution equation taking into account the
above boundary condition we get for the moment of $\qv^{\gamma}$:
\bea
&&\hspace{-1cm}
\qv^{\gamma}(n,t)=\frac{\alpha}{8\pi\beta_0}
\frac{4\pi}{\alpha_s(t)}
\Kv_n^{(0)}\sum_i P^n_i\frac{1}{1+d^n_i}
\left\{1-\left[\frac{\alpha_s(t)}{\alpha_s(0)}\right]
^{1+d^n_i}\right\}\nonumber\\
&&\hspace{2.5cm}+\qv^{\gamma}(n,0)
\sum_i P^n_i
\left[\frac{\alpha_s(t)}{\alpha_s(0)}\right]^{d^n_i}~,\label{formula}
\eea
where the $P^n_i$ is the projection operator onto
the eigenstate $\lambda_i$ of the anomalous dimension matrices
$\hat{\gamma}_n$:
\bea
\hat{\gamma}_n=\sum_iP^n_i\lambda^n_i~,
\eea
where the anomalous dimension matrices $\hat{\gamma}_n$ is 
related to the splitting function $P(x)$ as
\bea
\hat{\gamma}_n\equiv
-2\int_0^1 dx x^{n-1}P(x),\quad
\eea
and $d^n_i\equiv \lambda^n_i/2\beta_0$.
$\Kv^{(0)}_n$ is the anomalous dimension corresponding to
the photon-parton splitting function:
\bea
\Kv^{(0)}_n=2\int_0^1 dx x^{n-1}\kv^{0}(x)~.
\eea 
The initial value $\qv^\gamma(n,0)$ is determined so that we have
\bea
q_j^\gamma(t=t_{m_j})=0,\quad t_{m_j}=\frac{2}{\beta_0}
\ln\frac{\alpha_s(P^2)}{\alpha_s(m_j^2)}~,
\eea
or
\bea
&&\hspace{-1cm}
0=\frac{4\pi}{\alpha_s(t_{m_j})}
\sum_i \left(\Kv_n^{(0)}P^n_i\right)_j\frac{1}{1+d^n_i}
\left\{1-\left[\frac{\alpha_s(m_j^2)}{\alpha_s(P^2)}\right]
^{1+d^n_i}\right\}\nonumber\\
&&\hspace{1cm}+\sum_i
\left(\qv^\gamma(n,0)/\frac{\alpha}{8\pi\beta_0}
 P^n_i\right)_j
\left[\frac{\alpha_s(m_j^2)}
{\alpha_s(P^2)}\right]^{d^n_i}~,
\label{coupled-eq}
\eea
for $j=\lambda,s_{LS},q_H,s_H$ and $s_{LNS}$. 
By solving the above coupled equations we get
the initial condition:
$\qv^\gamma(n,0)=
(0,\hat{\lambda}(n),0,\hat{s}_{LS}(n),\hat{q}_H(n),
\hat{s}_H(n),0,\hat{s}_{LNS}(n))$. 

Now we write down the moments of the structure function
in terms of the parton distribution functions and
the coefficient functions, which are ${\cal O}(\alpha_s^0)$
at LO.
We take
\bea
\Cv^{(0)}_n(1,0)^T=
(0,0,\langle e^2\rangle_L, \langle e^2\rangle_L,e_H^2,e_H^2,1,1)~.
\eea
Then the $n$-th moment of the 
structure function $F_2^\gamma$ to the leading order in SQCD
is given by
\bea
&&\hspace{-1cm}
M_n^\gamma=\int_0^1 dx x^{n-1} F_2^\gamma/x
=\qv^\gamma(n)\cdot\Cv^\gamma_n(1,0)\nn\\
&&\hspace{-1cm}
=\langle e^2\rangle_L q_{LS}+\langle e^2\rangle_L s_{LS}
+e_H^2 q_H^2+ e_H^2 s_H+q_{LNS}+s_{LNS}~.\label{F2moment}
\eea

%%%%%%%%%%%%%%%%%%%%%%%%%%%%%%
\section{Numerical analysis}
%%%%%%%%%%%%%%%%%%%%%%%%%%%%%%

We have solved the equations (\ref{coupled-eq}) for $\qv^\gamma(n,0)$
numerically, and plug them into the master formula  (\ref{formula})
for the parton distribution functions and then evaluate the moments of 
the structure function $F_2^\gamma$ based on the formula (\ref{F2moment}). 
By inverting the Mellin moment we get the $F_2^\gamma$ as a function 
of Bjorken $x$.

%%%%%%%%%%%%%%%%%%%%%%%%%%%%%
%\vspace{2cm}
\begin{figure}[hbt]
\begin{center}
\includegraphics[scale=0.7]{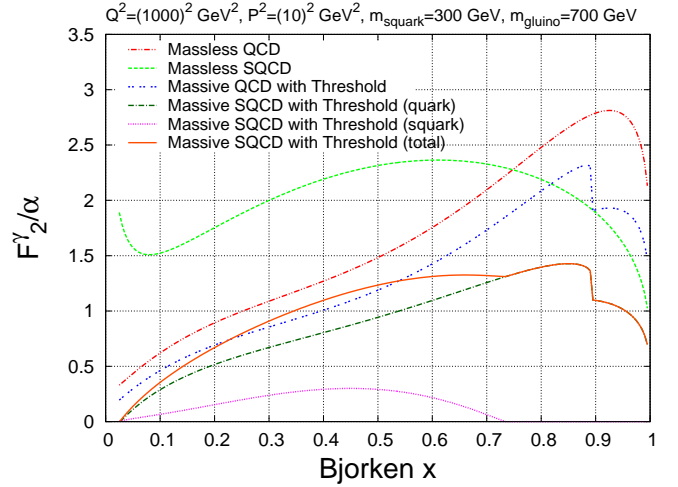}
\caption{$F_2^\gamma(x,Q^2,P^2)/{\alpha}$ with SUSY particles
as well as top threshold included. The dashed (2dot-dashed) curve
corresponds to the massless SQCD (QCD) case. The double-dotted
curve shows the massive QCD case. The dash-dotted (dotted)
curve corresponds to the quark (squark) component of the massive QCD. The
solid curve means the $F_2^\gamma/\alpha$ for the massive SQCD.
The kink at $x$=0.89 (0.74) corresponds to the top (squark) threshold.}
\end{center}
\end{figure}

In Fig. 3, we have plotted our numerical results for the
$F_2^\gamma/\alpha$. 
The 2dot-dashed and dashed curves correspond to the 
$F_2^\gamma/\alpha$ for the massless QCD and SQCD, 
respectively, where all the quarks and squarks are taken to 
be massless. Of course this is the unrealistic case we discussed in 
the previous section. For the more realistic case, we take $n_f=6$
and treat the $u$, $d$, $s$, $c$ and $b$ to be massless and take the 
top quark $t$ massive. We assume that all the squarks possess the same 
heavy mass and the gluino has another heavy mass.
In these analyses, we have taken $Q^2=(1000)^2$GeV$^2$ and
$P^2=(10)^2$GeV$^2$. For the mass values we took the top mass
$m_{\rm top}=175$ GeV, the common squark mass, $m_s=300$ GeV and
the gluino mass $m_{\lambda}=700$ GeV.

The double-dotted curve shows $F_2^\gamma/\alpha$ for the QCD 
with the mass of the top quark as well as the threshold effects taken 
into account. 
The dash-dotted curve shows the quark component for the massive SQCD
case with massive top quark, while the dotted curve means the squark
component for the same case. The sum of these leads to the solid curve 
which corresponds to $F_2^\gamma/\alpha$ for the massive 
SQCD with massive top and threshold effects included.
Here, we adopt the prescription for taking into account the threshold 
effects by rescaling the argument of the distribution function $f(x)$ 
as \cite{acot}:
\bea
f(x) \rightarrow f(x/x_{\rm max})~,\quad x_{\rm max}=
\frac{1}{1+\frac{P^2}{Q^2}+\frac{4m^2}{Q^2}}
\eea
where $x_{\rm max}$ is the maximal value for the Bjorken variable.
After this substitution the range of $x$ becomes $0\leq x\leq
x_{\rm max}$.
At small $x$, there is no significant difference between massless 
and massive QCD, while there exists a large difference 
between massless and massive SQCD. At large $x$, the significant 
mass-effects exist both for non-SUSY and SUSY QCD. The SQCD case
is seen to be much suppressed at large $x$ compared to the QCD.
The squark contribution to the total structure function in massive 
SQCD appears as a broad bump for $x<x_{\rm max}$. Here of course
we could set the squark mass larger than 300 GeV, {\it e.g.}
around 1 TeV, as recently reported by the ATLAS/CMS group at LHC,
for higher values of $Q^2$.

%%%%%%%%%%%%%%%%%%%%%%%%%%%%%%%%%%%%%%%
\section{Conclusion} \label{conclusion}
%%%%%%%%%%%%%%%%%%%%%%%%%%%%%%%%%%%%%%%

In this paper we have studied the virtual photon structure
function in the framework of the parton evolution equations
for the supersymmetric QCD, where we have PDFs for the squarks 
and gluinos in addition to those for the quarks and gluons. 
%We first studied the massless
%case and compared $F_2^\gamma$ for the massless QCD with
%that of the massless SQCD constructed out of the massless PDFs
%in QCD as well as in SQCD. 

%We then considered the heavy parton mass effects for the top 
We considered the heavy parton mass effects for the top quark, 
squarks and gluinos by imposing the boundary 
conditions for their PDFs in the framework treating heavy
particle distribution functions \cite{KSUUResum}. 
The PDF for the heavy particle with mass squared, $m^2$ 
are required to vanish at $Q^2=m^2$. 
This can be translated into the initial condition for the heavy 
parton PDFs, $\qv^\gamma(t=0)$. Due to the initial condition
the solution to the evolution equation is altered as given by 
(\ref{formula}). This change leads to the heavy mass effects for
the PDFs. As we have shown in Fig.3, there is no significant 
difference in the small-$x$ region between QCD and SQCD, 
while at large $x$, it turns out that there exists a sizable
difference between the massive QCD and SQCD. 
When compared to the squark contribution to $F_2^\gamma$ in the 
parton model calculation \cite{KYSU}, the squark component in 
the SQCD is suppressed at large $x$ due to the radiative correction. 
We expect that the future linear collider would enable such an 
analysis to be carried out on photon structure functions.
%--------- end of main text ---------

%---------------- Acknowledgments  -------------------%
%\section*{Acknowledgements}
%---------------- End of Acknowledgments  ------------%
%%%%%%%%%%%%%%%%%%%%%%%%%%%%%%%%%%%%%%%%%%%%%%%%%%%%%%%
%%%%%%%%%%%%%%%%%    Appendix  %%%%%%%%%%%%%%%%%%%%%%%%
%%%%%%%%%%%%%%%%%%%%%%%%%%%%%%%%%%%%%%%%%%%%%%%%%%%%%%%
%\newpage
\appendix
\section{Anomalous Dimensions for SUSY QCD}
Note that the our convention for the anomalous dimension
is related to the above splitting function as
\bea
\gamma_{ij}^n=-2\int_0^1 dx x^{n-1} P_{ij}(x)~.
\eea
The 1-loop anomalous dimensions for SUSY QCD are given by
\bea
&&\hspace{-1cm}
\gamma_{qq}^n=2C_F\left[-2-\frac{2}{n(n+1)}+4S_1(n)\right],
\nn\\
&&\hspace{-1cm}
\gamma_{sq}^n=2C_F\left[\frac{-2}{n+1}\right],\nn\\
&&\hspace{-1cm}
\gamma_{qs}^n=2C_F\left[\frac{-2}{n}\right],\nn\\
&&\hspace{-1cm}
\gamma_{ss}^n=2C_F\left[-2+4S_1(n)\right],\nn\\
&&\hspace{-1cm}
\gamma_{qG}^n=-4n_f\frac{n^2+n+2}{n(n+1)(n+2)},\nn\\
&&\hspace{-1cm}
\gamma_{sG}^n=-4n_f\frac{2}{(n+1)(n+2)}=-8n_f\frac{1}{(n+1)(n+2)},\nn\\
&&\hspace{-1cm}
\gamma_{q\lambda}^n=-4n_f\left[\frac{1}{n}-\frac{1}{n+1}\right],\nn\\
&&\hspace{-1cm}
\gamma_{s\lambda}^n=-4n_f\left(\frac{1}{n+1}\right),\nn\\
&&\hspace{-1cm}
\gamma_{Gq}^n=-4C_F\frac{n^2+n+2}{n(n^2-1)},\nn\\
&&\hspace{-1cm}
\gamma_{\lambda q}^n=-4C_F\frac{1}{n(n+1)},\nn\\
&&\hspace{-1cm}
\gamma_{Gs}^n=-4C_F\left[\frac{2}{n-1}-\frac{2}{n}\right],\nn\\
&&\hspace{-1cm}
\gamma_{\lambda s}^n=-4C_F\frac{1}{n},\nn\\
&&\hspace{-1cm}
\gamma_{GG}^n=2C_A\left[-3-\frac{4}{n(n-1)}-\frac{4}{(n+1)(n+2)}
+4S_1(n)\right]+2n_f, \nn\\
\nn\\
&&\hspace{-1cm}
\gamma_{\lambda G}=-4C_A\frac{n^2+n+2}{n(n+1)(n+2)}=
-12\frac{n^2+n+2}{n(n+1)(n+2)},\nn\\
&&\hspace{-1cm}
\gamma_{G\lambda}^n=-4C_A\left[\frac{2}{n-1}-\frac{2}{n}
+\frac{1}{n+1}\right], \nn\\
&&\hspace{-1cm}
\gamma_{\lambda \lambda}=2C_A\left[-3-\frac{2}{n}
+\frac{2}{n+1}+4S_1(n)\right],
\eea
%where $C_F=4/3$, $T_R=n_f$ and $C_A=3$ for SQCD.
where $C_F=4/3$ and $C_A=3$ for SQCD.
Hence we have the following anomalous dimensions 
for the supersymmetric case:
\bea
&&\hspace{-1cm}
\gamma_{\phi\phi}^n=\gamma_{qq}^n+\gamma_{sq}^n=
\gamma_{qs}^n+\gamma_{ss}^n
=2C_F\left[-2-\frac{2}{n}+4S_1(n)\right], \nn
\\
\nn\\
&&\hspace{-1cm}
\gamma_{\phi V}^n=\gamma_{qG}^n+\gamma_{sG}^n=
\gamma_{q\lambda}^n+\gamma_{s\lambda}^n
=-4n_f\frac{1}{n},\nn\\
&&\hspace{-1cm}
\gamma_{V\phi}^n=\gamma_{Gq}^n+\gamma_{\lambda q}^n=
\gamma_{Gs}^n+\gamma_{\lambda s}^n
=-4C_F\left[\frac{2}{n-1}-\frac{1}{n}\right],\nn\\
&&\hspace{-1cm}
\gamma_{VV}^n=\gamma_{GG}^n+\gamma_{\lambda G}^n=
\gamma_{G\lambda}^n+\gamma_{\lambda \lambda}^n\nn\\
&&=2C_A\left[-3-\frac{4}{n-1}+\frac{2}{n}
+4S_1(n)\right]+2n_f,
\eea
where we have the following replacement:
$n_f\gamma_{\phi V}^n\rightarrow \gamma_{\phi V}^n$.
In the case of non-supersymmetric QCD we have the
following anomalous dimensions:
\bea
&&\hspace{-1cm}
\gamma_{\psi\psi}^{0,n}=\gamma_{NS}^{0,n}
=\frac{8}{3}\left[-3 -\frac{2}{n(n+1)}+4S_1(n)\right],\nn\\
&&\hspace{-1cm}
\gamma_{\psi G}^{0,n}=-4n_f\frac{n^2+n+2}{n(n+1)(n+2)},\nn\\
&&\hspace{-1cm}
\gamma_{G\psi}^{0,n}=-\frac{16}{3}\frac{n^2+n+2}{n(n^2-1)},\nn\\
&&\hspace{-1cm}
\gamma_{GG}^{0,n}=6\left[-\frac{11}{3}-\frac{4}{n(n-1)}
-\frac{4}{(n+1)(n+2)}+4S_1(n)\right]\nn\\
&&\hspace{5cm}+\frac{4}{3}n_f.
\eea
%%%%%%%%%%%%%%%%%%%%%%%%%%%%%%
\section{Photon-parton mixing anomalous dimensions}
%%%%%%%%%%%%%%%%%%%%%%%%%%%%%%
The photon-parton splitting function can be connected
to the photon-parton mixing anomalous dimensions given by
\bea
\Kv^{(0)}_n=\left(K^{0,n}_G,K^{0,n}_\lambda,K^{0,n}_{q_{LS}},
K^{0,n}_{s_{LS}},
K^{0,n}_{q_H},K^{0,n}_{s_H},K^{0,n}_{q_{LNS}},K^{0,n}_{s_{LNS}}\right), \nn\\
\eea
where
\bea
&&K^{0,n}_G=K^{0,n}_\lambda=0,\nn\\
&&K^{0,n}_{q_{LS}}=24(n_f-1)\langle e^2\rangle_L
\frac{n^2+n+2}{n(n+1)(n+2)},\nn\\
&&K^{0,n}_{s_{LS}}=24(n_f-1)\langle e^2\rangle_L
\left[\frac{1}{n}-\frac{n^2+n+2}{n(n+1)(n+2)}\right],\nn\\
&&K^{0,n}_{q_H}=24e_H^2\frac{n^2+n+2}{n(n+1)(n+2)},\nn\\
&&K^{0,n}_{s_H}=24e_H^2
\left[\frac{1}{n}-\frac{n^2+n+2}{n(n+1)(n+2)}\right],\nn\\
&&K^{0,n}_{q_{LNS}}=24(n_f-1)\left(\langle e^4\rangle_L
-\langle e^2\rangle_L^2\right)\frac{n^2+n+2}{n(n+1)(n+2)},\nn\\
&&K^{0,n}_{s_{LNS}}=24(n_f-1)\left(\langle e^4\rangle_L
-\langle e^2\rangle_L^2\right)
\left[\frac{1}{n}-\frac{n^2+n+2}{n(n+1)(n+2)}\right].\nn\\
\eea
%%%%%%%%%%%%%%%%%%%%%%%%%%%%%%%%%%%%%%%%%%%

%--------- End of Appendix ----------

%---------- References -------------

%\bibliographystyle{elsarticle-num}



\end{document}